# Work-loop techniques for optimising nonlinear forced oscillators

Arion Pons [*][**][†], and Tsevi Beatus [*][**][‡]
[*]*Institute of Life Sciences, Hebrew University of Jerusalem, Israel*
[**]*School of Computer Science and Engineering, Hebrew University of Jerusalem, Israel*

*Summary*. Linear and nonlinear resonant states can be restrictive: they exist at particular discrete states in frequency and/or elasticity, under particular (*e.g.*, simple-harmonic) waveforms. In forced oscillators, this restrictiveness is an obstacle to system design and control modulation: altering the system elasticity, or modulating the response, would both appear to necessarily incur a penalty to efficiency. In this work, we describe an approach for bypassing this obstacle. Using novel work-loop techniques, we prove and illustrate how certain classes of resonant optimisation problem lead to non-unique solutions. In a structural optimisation context, several categories of energetically-optimal elasticity are non-unique. In an optimal control context, several categories of energetically-optimal frequency are non-unique. For these classes of non-unique optimum, we can derive simple bounds defining the optimal region. These novel theoretical results have practical implications for the design and control of a range of biomimetic propulsion systems, including flapping-wing micro-air-vehicles: using these results, we can generate efficient forms of wingbeat modulation for flight control.

## 1. Introduction

A wide range of biological and engineering systems utilize nonlinear structural elasticity to shape and control an oscillatory forced response. Insects rely on strain-hardening thoracic elasticity to modulate flight motor muscular actuation [1, 2]. Micro-mechanical energy harvesters utilise nonlinear elasticity to improve harvesting efficiency [3–5]. Compliant bipedal robots utilise distributed elasticity to improve walking efficiency [6]. Among the multifaceted roles that nonlinear elasticity can play in such systems, energetic roles are often central: structural elasticity can absorb inertial loads and/or mitigate inertial power requirements, thereby increasing system efficiency. However, there is a sense in which the pursuit of efficiency can be restrictive. Energetically-optimal states tend to be discrete: located at particular states of elasticity (*e.g.*, the resonant elasticity); or at particular discrete frequencies (*e.g.*, the resonant frequencies), and particular (*e.g.*, symmetric, harmonic) waveforms. For instance: as per classical analysis, linear resonance occurs at discrete frequencies, and under simple-harmonic forcing; and in nonlinear systems, such as the Duffing oscillator, frequency response magnitude peaks are also typically discrete. In both cases, deviating from the energetically-optimal frequency-elasticity match, or resonant state, incurs a penalty in efficiency.

Here, we describe a novel technique for bypassing this efficiency penalty in some contexts. Using work-loop analysis techniques – showing several parallels with phase portrait techniques – we can illustrate and prove several key optimality results. These techniques allow us to prove how several classes of forward- and inverse-problems for energetic optimality necessarily lead to non-unique solutions; and to derive simple bounds defining this region of non-unique optimal solutions. These new theoretical results have significant implications for the use of nonlinear dynamics in engineering design. They describe ways to introduce nonlinear elasticities into a system to ensure energetic optimality, and illustrate how, for certain classes of energetic optimality, a whole space of optimal nonlinear elasticities exists. The choice of nonlinear elasticity within this space is a design tool that can be used to control other aspects of system behaviour. These results also describe methods for modulating the frequency of a resonant response, and/or breaking its symmetry, while maintaining energetic optimality. Improving the energetic optimality of these forms of response modulation is crucial to several forms of bio-inspired locomotion system: modulating the frequency of bipedal walking can govern the transition to running [7]; and modulating the wingstroke offset of a flapping-wing micro-air-vehicle (FW-MAV) can lead to body pitch control [8]. In this way, these theoretical techniques lead to new design and control principles for a range of engineering systems.

## 2. Energy resonance in the time-domain

In linear and nonlinear systems, the phenomenon of resonance is complex and multifaceted: representing a range of distinct, and often mutually-exclusive, states of optimality [9, 10] and interaction phenomena [11]. Energy, or global, resonance is one such resonant phenomenon that has key relevance to the design and operation of efficient resonators (biolocomotive systems, energy harvesters, *etc.*) [3–5, 9, 12]. Consider, for instance, a general nonlinear time-invariant single-degree-of-freedom (1DOF) parallel elastic actuation (PEA) system (Fig. 1):

$$D(x, \dot{x}, \ddot{x}, \dots) + F_s(x) = F(t). \qquad (1)$$

---

[†] arion.pons@mail.huji.ac.il
[‡] tsevi.beatus@mail.huji.ac.il

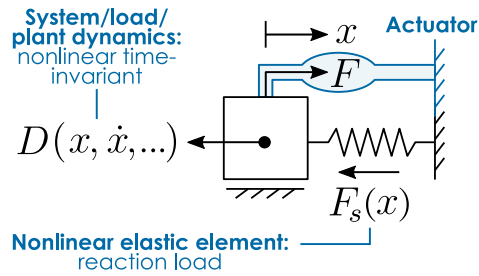

Figure 1: Schematic of a general nonlinear time-invariant 1DOF PEA system

with general system (*i.e.*, plant, or load) dynamics $D(x, \dot{x}, \ddot{x}, \ldots)$, linear or nonlinear elastic element $F_s(x)$, and actuator load input $G(t)$. The state of energy resonance in this system is easy to define, but hard to locate – it is the state:

$$F(t)\dot{x}(t) \geq 0, \forall t. \qquad (2)$$

That is, the actuator power consumption should always be positive: power should flow from the actuator to the system, and never vice-versa. This unidirectionality of power flow represents both an intuitive state of energetic optimality, and a formal one. Intuitively, if we desire the largest possible system response, then we should seek always to use the actuator, $F(t)$, to feed energy into the system – rather than draw energy out of the system. Unidirectional power flow defines the condition that all of the actuator power consumption is directed towards increasing the total energy in the system.

Formally, energy resonant states are optimal with respect to elasticity. Consider an energy resonant state for Eq. 1 at some $F_s(x)$, and some $x(t)$ and $F(t)$ that are periodic with common period $T$. The overall actuator mechanical power consumption at this state can be defined in multiple ways, for instance:

$$\text{the net power:} \qquad \overline{P}_{\text{net}} = \int_0^T F(t)\dot{x}(t)\, dt,$$
$$\text{the absolute power:} \qquad \overline{P}_{\text{abs}} = \int_0^T |F(t)\dot{x}(t)|\, dt, \qquad (3)$$
$$\text{the positive-only power:} \qquad \overline{P}_{\text{pos}} = \int_0^T F(t)\dot{x}(t)[F(t)\dot{x}(t) > 0]_I\, dt,$$

where $[\,\cdot\,]_I$ is the Iverson bracket, $[\lambda]_I = 1$ for $\lambda$ true, $[\lambda]_I = 0$ for $\lambda$ false [13]. This difference between these metrics lies in their treatment of negative power, $F\dot{x} < 0$. The net power represents the mechanical power throughput of the system – the power dissipated to the system or plant, $D(\cdot)$. It does not represent the mechanical power consumption of the actuation except in the case where the actuator is intrinsically capable of absorbing negative power, and storing this power for future use – for instance, in certain configurations of electrical actuator with power electronics. The absolute power represents the mechanical power consumption of the actuator if the actuator itself is responsible for drawing power out of the system – for instance, a rocket or jet engine, which must consume fuel to produce thrust, irrespective of the direction this thrust is oriented (whether to produce negative or positive power). The positive-only power represents the mechanical power consumption of the actuator if an additional dissipative braking system is responsible for generating negative power – for instance, in a vehicle with a dissipative braking system. A dissipative braking system can relieve the actuator from a responsibility to generate negative power, but cannot store this energy for future use.

Note that further, more generalised, negative-power penalties can be defined, *e.g.*, to represent the behaviours of biological muscles [12, 14, 15]. However, the penalty detail matters not: it is clear that, only under the energy resonant condition, Eq. 2, will the power throughput (net power) equal the actuator power consumption (absolute, positive-only, *etc.*). In any non-energy resonant state, the actuator power consumption will be greater than the net power; the difference in power representing wasted negative power. A key corollary of this is that, at an energy resonant state, there exists no other elasticity, $F_s(x)$, that could reduce the actuator power consumption (absolute, positive-only, *etc.*) required to generate $x(t)$. This can be demonstrated in the following way. The elasticity, $F_s(x)$, cannot alter the net power required to generate a specified $x(t)$ – this can be confirmed by evaluating the net power integral:

$$\int_0^T F(t)\dot{x}(t)\, dt = \int_0^T D(x, \dot{x}, \ddot{x}, \ldots)\dot{x}(t)\, dt + \int_{x(0)}^{x(T)} F_s(x)\, dx = \int_0^T D(x, \dot{x}, \ddot{x}, \ldots)\dot{x}(t)\, dt. \qquad (4)$$

It is also impossible for the actuator power consumption to be less than the net power – this is a property of the definitions in Eq. 3. Therefore, there exists no elasticity $F_s(x)$, that could further minimise the actuator power consumption (absolute, positive-only, *etc.*) required to generate the output, $x(t)$, of an energy resonant state. In this sense, energy resonant states are energetically-optimal.

### 3. Energy resonance in work-loop planes

Work loops are an analysis tool used frequently in applied contexts, but have not yet seen significant theoretical application. Their defining characteristic is that they display a metric of force against a metric displacement, and therefore have an area that is synonymous with work, *i.e.*, energy. In biomechanics, work loops are used to visualise, and characterise the behaviour of muscle groups undergoing periodic motion – *e.g.*, muscles within the insect flight motor [16–18]. In materials science, they are one of many forms of hysteresis loop, used to characterise viscoelastic and other hysteretic material behaviour [19]. Work-loop techniques can be applied to the dynamical system of Eq. 1: given some pair of $x(t)$ and $F(t)$ for Eq. 1, it is possible to visualise this pair in the plane of $F$ against $x$ (Fig. 2). We denote this plane $x$-$F$ (abscissa-ordinate). If $x(t)$ and $F(t)$ are periodic, with common period, then this pair will trace out a closed loop in the $x$-$F$ place: a work loop. This work loop may take many shapes. For instance, any linear PEA system undergoing steady-state simple-harmonic motion shows an elliptical work loop (Fig. 2a). That is, for a system of the form:

$$\ddot{x} + 2\zeta\omega_0\dot{x} + \omega_0^2 x = F(t), \tag{5}$$

with $x = \hat{x}\cos(\Omega t)$, we can establish that:

$$F^2 - 2F(\omega_0^2 - \Omega^2)x + ((\omega_0^2 - \Omega^2)^2 + 4\zeta^2\omega_0^2\Omega^2)x^2 = 4\zeta^2\omega_0^2\Omega^2\hat{x}^2, \tag{6}$$

describing an ellipse. However, categories of dynamical system do not directly translate to consistent work loop shapes. For instance, as we add additional harmonics to the motion of a linear system, the work loop rapidly becomes inexpressible in closed form.

In our analysis, we will consider a system which is simultaneously more general and more restrictive than the nonlinear PEA system of Eq. 1. We consider a work loop that is a closed simple curve (*i.e.*, no self-intersections), and is no more than bivalued at any $x$ (Fig. 2). Other than this, the shape of the loop can be arbitrary. Such a loop could arise from systems considerably more complex than Eq. 1, including, *e.g.*, computational fluid dynamics (CFD) models. For instance, Fig. 2b illustrates a work loop arising from CFD analysis of *Drosophila melanogaster* wingbeat oscillation, derived from [20–22]. But, conversely, even the linear PEA system, Eq. 5, is capable of generating work loops which are more than bivalued at any $x$ – for instance, very simply, in cases where the output, $x(t)$, is not composed of two monotonic half cycles.

In cases where the work loop is a closed simple curve, as specified, it is representable as an upper and lower curve: $F^+(x)$ and $F^-(x)$ (Fig. 2a). These curves represent the two monotonic half-cycles of $x(t)$, and, as such, are each associated with a particular sign of the velocity ($\dot{x}$). For a dissipative work loop (net power $> 0$), over $F^+$, $\dot{x} > 0$, and over $F^-$, $\dot{x} < 0$. Going further, if we distinguish between the system's elastic load, $F(t)$, and inelastic load, $G(t)$, as altered by a parallel (PEA) elasticity – that is, in the particular case of Eq. 1, as:

$$\begin{aligned}\text{inelastic load, } G(t): \quad & D(x, \dot{x}, \ddot{x}, \ldots) = G(t), \\ \text{elastic load, } F(t): \quad & D(x, \dot{x}, \ddot{x}, \ldots) + F_s(x) = G(t) + F_s(x) = F(t),\end{aligned} \tag{7}$$

– then we can define work-loop equations of motion for this general PEA system:

$$G^\pm(x) + F_s(x) = F^\pm(x). \tag{8}$$

These work-loop equations of motion describe not only Eq. 1, but also the load-requirement dynamics of a more complex system (*e.g.*, the CFD model in Fig. 2b): elasticity, $F_s(x)$ alters the load, $F^\pm(x)$, required to generate the output response associated with the desired output, $G^\pm(x)$. In simple cases, Eq. 7-8 can be expressed entirely in closed form. For instance, the linear PEA system undergoing simple-harmonic motion (Eq. 5):

$$\begin{aligned}G^\pm(x) &= \Omega^2 x \pm 2\zeta\omega_0\Omega\sqrt{\hat{x}^2 - x^2} \\ F^\pm(x) &= (\omega_0^2 - \Omega^2)x \pm 2\zeta\omega_0\Omega\sqrt{\hat{x}^2 - x^2}\end{aligned} \tag{9}$$

$$i.e., G^\pm(x) + \omega_0^2 x = F^\pm(x).$$

The work-loop equations of motion, Eq. 8, permit a definition of the energy resonant condition, Eq. 2, in the work-loop plane. We seek to define the condition $F\dot{x} \geq 0$ (Eq. 2), and we know that over $F^+$, $\dot{x} > 0$, and over $F^-$, $\dot{x} < 0$, therefore, for energy resonance:

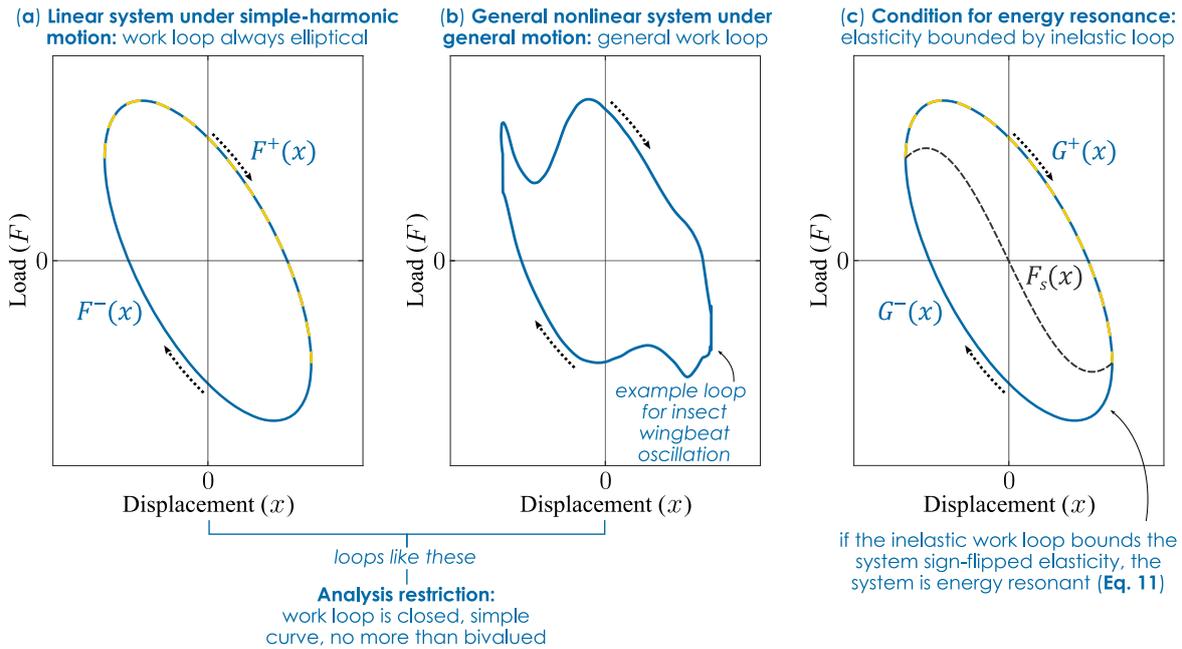

Figure 2: Schematic of work loops for (**a**) linear and (**b**) general systems, alongside (**c**) the conditions for energy resonance, in $F_s(x)$ and $G^\pm(x)$.

$$F^+(x) \geq 0 \text{ and } F^-(x) \leq 0. \tag{10}$$

Or, via Eq. 8, for energy resonance:

$$G^-(x) \leq -F_s(x) \leq G^+(x), \tag{11}$$

as illustrated in Fig. 2c. Eq. 11 is the elastic-bound condition for this general PEA system, describing the relationship between elasticity, $F_s(x)$, and the inelastic system work loop, $G^\pm(x)$ (a function only of the plant dynamics and desired output) that must exist in order for an energy resonant state to exist. Eq. 11 is an inequality condition, describing a continuum of states – whether in $F_s(x)$, or $G^\pm(x)$ – that are energy resonant. Notably, however, this condition contains within it a pair of equality condition. At $\max x$ and $\min x$ ($\pm \hat{x}$ for simple-harmonic motion), $G^-(x) = G^+(x)$, by virtue of the fact that the work loop is closed. Therefore, at these values, the inequality in Eq. 12 becomes an equality:

$$\begin{aligned} G^-(\max x) &= -F_s(\max x) = G^+(\max x), \\ G^-(\min x) &= -F_s(\min x) = G^+(\min x). \end{aligned} \tag{12}$$

Physically, peak inertial loads (inelastic loads at the $x$-extrema, where $\dot{x} = 0$) must always match peak elastic loads (elastic loads at the $x$-extrema). This equality condition, Eq. 12, is a useful commonality between the continuum of energy resonant states defined by Eq. 11 – we will illustrate its utility in Sections 4-5.

### 4. Work-loop analysis for structural optimisation

*4.1. Elastic-bound optimisation principle*
Practical applications of work-loop analysis arise in a range of physical and analytical contexts. One such analytical context is that of structural optimisation in nonlinear oscillators – for instance, the optimisation of oscillator structural properties so as to generate a desired forced oscillatory response at maximum efficiency. Structural optimisation problems of this form arise in a range of physically-relevant oscillatory systems: for instance, flapping-wing micro-air-vehicles, involving the design of drivetrain elastic elements to ensure maximum efficiency [23–25]; and micro-mechanical energy harvesters, involving the design of oscillators for maximum energy absorption [3–5]. The elastic-bound conditions, Eq. 11, are directly applicable in a structural optimisation context. Consider a nonlinear oscillator, of the form $D(x, \dot{x}, \ddot{x}, \dots) + F_s(x) = G(t)$ (Eq. 8), with some structural elasticity, $F_s(x)$, that we have design control over. If we specify some desired system output, a periodic $x(t)$, then, given some $F_s(x)$, we know the actuator load, a periodic $F(t)$, that is required to generate this desired output. The optimisation problem then is to select $F_s(x)$ such that this load requirement, $F(t)$, is optimal in some way. If we are interested in energy-efficiency, then a key metric to optimise is the mechanical power consumption associated with $F(t)$: for instance, the absolute power, $\overline{P}_{abs}$, or positive-only power, $\overline{P}_{pos}$, as per Eq. 3[§]. In either case, the elastic-bound conditions provide a direct solution to this optimisation problem: it is the elasticities, $F_s(x)$, satisfying Eq. 11, that ensure that $\overline{P}_{abs}$ and $\overline{P}_{pos}$ take their minimum values with respect to $F_s(x)$. This is a highly-general principle, applicable to any PEA system that generates a closed, simple, bi-valued work loop, under the specified $x(t)$ (*cf.* Section 3). Fig. 3a illustrates this principle, alongside the two more specific cases studied below.

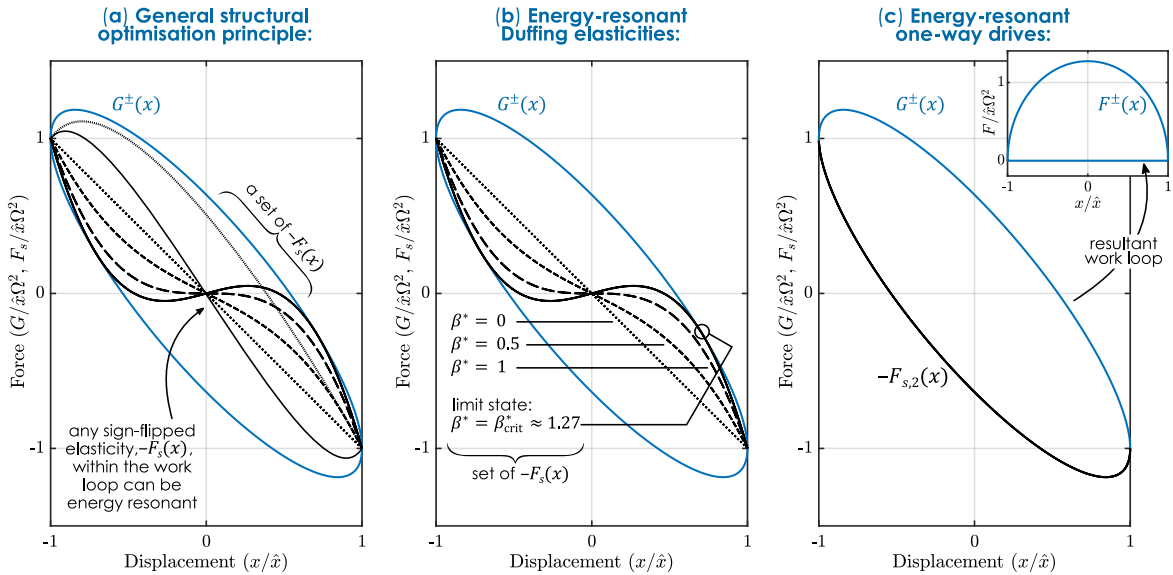

Figure 3: Work loop analysis and energy resonance in a structural optimization context. (**a**) The general structural optimisation principle (Section 4.2); (**b**) a set of energy-resonant Duffing elasticities (Section 4.2); (**c**) and energy-resonant one-way-drive elasticity (Section 4.3). The Duffing oscillator inelastic parameters are $\delta = 4$, $\omega = 2\pi$, $\hat{x} = 1$.

*4.2 Structural optimisation of a Duffing oscillator*
As an example of the work-loop structural optimisation process, consider the Duffing oscillator:

$$\ddot{x} + \delta\dot{x} + \alpha x + \beta x^3 = F(t) \tag{13}$$

with damping $\delta > 0$, linear stiffness $\alpha$, cubic stiffness $\beta$, and input loading $F(t)$ which, in our analysis, may not necessarily be simple-harmonic. The energy-minimisation structural optimisation problem in this oscillator is to compute the elastic parameters, $\alpha$ and $\beta$, such that this specified output can be generated at a state of energy resonance, by some $F(t)$. As such, in a work-loop context, we can split Eq. 13 into an inelastic load, $G(t)$, and an elastic profile, $F_s(x)$:

$$\ddot{x} + \delta\dot{x} = G(t), \qquad F_s(x) = \alpha x + \beta x^3 \tag{14}$$

More specifically, consider a steady-state simple-harmonic output, $x = \hat{x}\cos(\Omega t)$. In this case, the inelastic work loop is given by the elliptical profile:

$$G^{\pm}(x) = -\Omega^2 x \pm \delta\Omega\sqrt{\hat{x}^2 - x^2}. \tag{15}$$

The elastic-bound equality conditions (Eq. 12) translate to conditions on the stiffness parameters, $\alpha$ and $\beta$:

$$-F_s(\hat{x}) = G^{\pm}(\hat{x}) \therefore \Omega^2\hat{x} = \alpha\hat{x} + \beta\hat{x}^3 \therefore \alpha = \Omega^2 - \beta\hat{x}^2 \tag{16}$$

which is to say, the Duffing stiffnesses, $F_s(x)$ that can generate this simple-harmonic output at a state of energy resonance are necessarily of the form

$$\frac{F_s(x)}{\Omega^2} = (1 - \beta^*\hat{x}^2)x + \beta^*x^3 \tag{17}$$

where $\beta^* = \beta/\Omega^2$ is a free parameter. Note that the equality condition, yielding Eq. 17, is necessary for the existence of an energy resonant state, but no sufficient. We anticipate, based on the full elastic-bound conditions, that Eq. 17 should describe an energy-resonant state only over some range of $\beta^*$. We may compute this range via the full elastic-bound conditions. Consider that, at the point at which the elastic-bound conditions fail, there exists the critical state:

$$G^+(x) = F_s(x) \text{ or } G^-(x) = F_s(x), \text{ for some } x \neq \pm\hat{x}. \tag{18}$$

---

[§] Note that, even if the net power, $\overline{P}_{net}$, were to be relevant to the actuator under consideration, $\overline{P}_{net}$ is independent of $F_s(x)$: the structural optimisation problem associated with $\overline{P}_{net}$ is irrelevant. This is further illustration of how $\overline{P}_{abs}$ and $\overline{P}_{pos}$ are more relevant metrics of actuator mechanical power consumption.

That this state necessarily exists is a property of the smoothness of this system. Again, with the elliptical inelastic work loop profile, Eq. 15, we can describe $x$ such that Eq. 18 is satisfied, as the roots of the quartic polynomial:

$$x^4 - x^2\hat{x}^2 + \frac{\delta^2}{\beta^{*2}\Omega^2} = 0. \tag{19}$$

Eq. 19 is a quadratic polynomial in $x^2$, and, as such, we can determine that the critical $x$ will exist (as a real number) if and only if the discriminant of this quadratic is positive. This leads to the condition:

$$-\beta^*_{\text{crit}} \leq \beta^* \leq \beta^*_{\text{crit}}, \qquad \beta^*_{\text{crit}} = \frac{2\delta}{\hat{x}^2\Omega} \tag{20}$$

For the constrained Duffing elasticity, Eq. 19, to be capable of generating an energy-resonant simple-harmonic output, under some periodic forcing. The actual periodic forcing required is trivial to define – it is given by Eq. 13 – though note that chaotic effects may necessitate a control system to maintain a steady-state output (and may even mean that such control is impractical). Nevertheless, Eq. 17 and 20 define the Duffing stiffnesses that are required for an energetically-optimal simple-harmonic output. Other non-simple-harmonic outputs can be optimised via the same work-loop analysis process, though numerical techniques may be required to compute both the form of the optimal elasticity (Eq. 17) and the range over which it is valid (Eq. 20)

*4.3 Special cases of energy-resonant structure*
In Section 4.2, we considered the optimisation of an elastic element of prescribed structure: the cubic elasticity of the Duffing oscillator. However, in a broader systems-design context, there may be no particular reason to restrict oneself to prescribed structures – considering elasticities with particular energy-resonant properties, irrespective of structure, may also be worthwhile. Work-loop analysis is a tool for identifying energy-resonant elasticities with special properties. As a notable example of this, consider again the elastic-bound conditions, Eq. 11. These conditions include, as an extremal case, the elasticities lying on the upper and lower boundaries of the loop:

$$\begin{aligned} F_{s,1}(x) &= -G^+(x), \\ F_{s,2}(x) &= -G^-(x). \end{aligned} \tag{21}$$

By definition, these elasticities generate a resultant work loop, $F^\pm(x)$, that has one half-cycle at zero load (Eq. 9, Fig. 3c):

$$\begin{aligned} F_{s,1}(x) = -G^+(x) \therefore F^-(x) = G^-(x) - G^+(x), \ F^+(x) = 0, \\ F_{s,2}(x) = -G^-(x) \therefore F^+(x) = G^+(x) - G^-(x), \ F^-(x) = 0. \end{aligned} \tag{22}$$

It follows also, that the other half-cycle will be at a state of unidirectional load: $G^+(x) - G^-(x)$ can never change sign, if $G^+(x) \geq G^-(x)$ as per our condition for no self-intersection of the work loop. The result is an energy-resonant state requiring only unidirectional actuation: we term this, a one-way-drive state. An additional property of this state is that it the duty cycle of the actuator is reduced: over the region of zero load requirement, the actuator need not be activated or energised. With the proper selection of one-way-drive elasticity, the actuator duty cycle can be reduced to 50% or less, for any appropriate work loop [12]. Both these properties make one-way-drive states systems ideal for certain classes of actuator: those which are capable of generating large intermittent loads in a single direction – such as combustion cylinders, explosive actuators, and solenoid actuators [26, 27]. Combustion cylinders, in particular, are more energy-dense than many forms of electromechanical actuator: utilising combustion cylinders to generate one-way-drive energy resonant oscillations in a flapping-wing aircraft may be an avenue to increasing the range and endurance of these aircraft.

## 5. Work-loop analysis for optimal control

*5.1. Elastic-bound optimal control principle*
Another practical application of work-loop analysis arises in the context of optimal control. In contrast with structural optimisation, which deals variable system structural properties, optimal control problems typically relate to a fixed system structure, and consider variable responses within this structure – with the aim to locate output responses (and, input forces) that are optimal in some sense. Work loop analysis is particularly relevant to optimal control problems that are concerned with optimising power consumption, or energetic efficiency, as it allows an easy description, and visualisation, of the location and properties of energy-resonant states. One of the interesting properties of energy resonance is that, even in the simplest linear systems, energy resonance is a non-unique state. As per Section 4, the energy-resonant structural elasticity for a system undergoing a specified output is non-unique

– there exists a continuum of energy-resonant elasticities. So also, the energy-resonant output for a system with fixed elasticity is non-unique – there exists a continuum of energy-resonant outputs. This continuum of energy-resonant outputs is defined directly by the elastic-bound conditions: for fixed elasticity $F_s(x)$, we seek $G^\pm(x)$ such that Eq. 11 is satisfied – $G^\pm(x)$ depending only on the desired output, $x(t)$, and the inelastic dynamics of the system, $D(\cdot)$**. We note that these different work loops may represent outputs at different amplitudes, frequencies, and symmetric or asymmetric waveforms. If then, as part of our control problem, we desire to modulate the frequency of our system output response, while maintaining a state of energy resonance (*i.e.*, optimality in mechanical power consumption), then work-loop analysis techniques provide a strategy for doing so. We refer to this energy-resonant frequency modulation as frequency-band resonance, and note that other forms of energy resonant modulation (*e.g.*, offset-band resonance, involving symmetry-breaking modulation [12]) are also available.

### 5.2. Optimal frequency control of a Duffing oscillator

Consider again the general Duffing oscillator, Eq. 14. As per our structural optimisation results (Eq. 16), the frequency, $\omega_e$, at which a simple-harmonic output is energy resonant in this general system is:

$$\omega_e^2 = \alpha + \beta \hat{x}^2. \tag{23}$$

As per our optimal control analysis, the question arises: can one maintain energy resonance while deviating from this energy-resonant frequency? As one instance of a deviation from a simple-harmonic output, consider the multiharmonic output given by:

$$\begin{aligned} x(t) &= \hat{x}(1-\rho)\cos(\Omega t) + \hat{x}\rho\cos(3\Omega t), \\ \dot{x}(t) &= -\Omega\hat{x}(1-\rho)\sin(\Omega t) - 3\Omega\hat{x}\rho\sin(3\Omega t), \\ \ddot{x}(t) &= -\Omega^2\hat{x}(1-\rho)\cos(\Omega t) - 9\Omega^2\hat{x}\rho\cos(3\Omega t), \end{aligned} \tag{24}$$

for small $\rho$, this multiharmonic wave is qualitatively sinusoidal. More precisely: for the window $-0.125 \leq \rho \leq 1$, computed numerically, the global displacement extrema of $x(t)$ are $\pm\hat{x}$ (max $x = \hat{x}$ and min $x = -\hat{x}$), and are located at $t = 0, T/2, T, \ldots$, where $T = 2\pi/\Omega$. For the more restrictive window $-0.125 \leq \rho \leq 0.25$, also computed numerically, $x(t)$ is composed of two monotonic half-cycles (velocity reversal, where $\dot{x} = 0$, exists only at $t = 0, T/2, T, \ldots$). We require both these conditions to be true – thus limiting ourselves to the more restrictive of these windows.

As part of a frequency-band analysis, we seek $\rho(\Omega)$ such that $x(t)$ is energy resonant at $\Omega$. Note that, at the energy-resonant frequency, $\omega_e$, a simple-harmonic wave is energy resonant (Section 4.2), and thus $\rho(\omega_e) = 0$. To compute the rest of $\rho(\Omega)$, note that the elastic-bound conditions necessitate:

$$G^\pm(\hat{x}) = F_s(\hat{x}), \text{ and } G^\pm(-\hat{x}) = F_s(-\hat{x}), \tag{25}$$

where we have assumed that max $x = \hat{x}$ and min $x = -\hat{x}$; an assumption we know to be valid over $-0.125 \leq \rho \leq 1$. Over this window we know, in addition, that these extrema will be located at $= 0, T/2, T, \ldots$, where $T = 2\pi/\Omega$. We can therefore translate Eq. 25 into a condition in the time-domain:

$$\begin{aligned} G(0) &= G(T) = F_s(\hat{x}) = \alpha\hat{x} + \beta\hat{x}^3, \\ G(T/2) &= F_s(-\hat{x}) = -\alpha\hat{x} - \beta\hat{x}^3, \end{aligned} \tag{26}$$

which, combining with load profile:

$$\begin{aligned} G(t) = \ddot{x} + \delta\dot{x} &= -\Omega^2\hat{x}(1-\rho)\cos(\Omega t) - 9\Omega^2\hat{x}\rho\cos(3\Omega t) \\ &\quad - \delta\Omega\hat{x}(1-\rho)\sin(\Omega t) - 3\delta\Omega\hat{x}\rho\sin(3\Omega t), \end{aligned} \tag{27}$$

yields the condition:

$$G(0) = G(T) = -G(T/2) = -\Omega^2\hat{x}(1+8\rho) = -F_s(\hat{x}) = -\alpha\hat{x} - \beta\hat{x}^3. \tag{28}$$

From Eq. 28 we deduce that the relationship $\rho(\Omega)$ is:

$$\rho(\Omega) = \frac{1}{8}\left(\frac{\alpha + \beta\hat{x}^2}{\Omega^2} - 1\right) = \frac{1}{8}\left(\frac{\omega_e^2}{\Omega^2} - 1\right). \tag{29}$$

This relationship ensures that the elastic-bound equality conditions (Eq. 12) will be satisfied in the Duffing oscillator. Again, we expect that this relationship will satisfy the full elastic-bound conditions (Eq. 11) only over a certain window of $\rho$ and $\Omega$. The properties of this window are determined by the Duffing parameters $\alpha, \beta, \delta$, and $\hat{x}$, but they are difficult to define analytically. The critical states defining the boundaries of this window are given as the solution of:

$$G(t) = -F_s(x), \qquad (30)$$

with $x(t)$ as per Eq. 24, and $\rho(\Omega)$ as per Eq. 29. We compute these critical states numerically. For instance, for $\alpha = 1$, $\beta = 3$, $\delta = 2$, and $\hat{x} = 1$, we estimate, conservatively, that frequency-band resonant states exist within $0.693\omega_e < \Omega < 1.163\omega_e$, for which $-0.0650 < \rho < 0.0441$. This range is illustrated in Fig. 4. This demonstration of frequency-band resonance raises the possibility of energetically-optimal frequency modulation in a range of physical systems – we refer, in particular, to insects and FW-MAVs, for which wingbeat frequency modulation can be an important form of lift and thrust control. This illustrates the phenomenon that energetically-optimal (*i.e.*, energy resonant) operation of linear and nonlinear systems is not restricted to a single resonant frequency, but is, in fact, available over a window of frequencies. Not that this effect is not a result of nonlinearity (it exists, for instance, when $\beta = 0$), but is a result of the damping in the system ($\delta$: the width of the work loop in the elastic-bound conditions, *cf.* Eq. 15). Damping allows energy-resonant behaviour over a range of frequencies – a remarkable physical phenomenon.

---

** Note that this is the same as saying that we seek $F^\pm(x)$ such that Eq. 10 is satisfied. This latter formulation does not necessitate that the system be split into 'elastic' and 'inelastic' components.

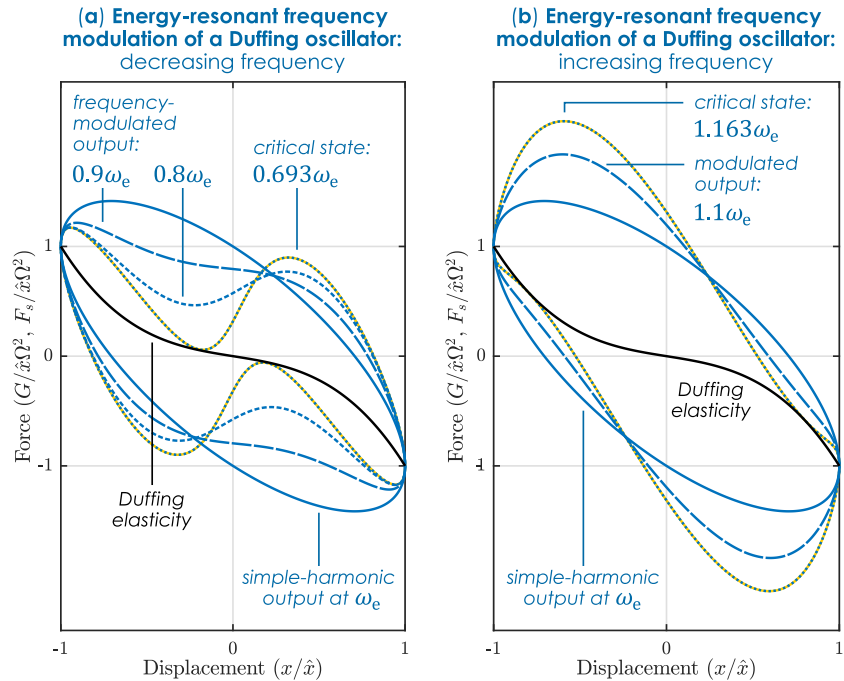

Figure 4: Work loop analysis and energy resonance in an optimal control context: frequency modulation of a Duffing oscillator, involving (**a**) decreasing and (**b**) increasing output frequency.

## 7. Conclusion

In this work we present several novel techniques of work-loop analysis, and demonstrate their theoretical and practical significance. Utilising them, we prove new optimality results for a wide class of nonlinear system, with implications for optimal system design and optimal control modulation in a range of engineering systems, including FW-MAVs. These optimality results have a range of implications in structural optimisation and optimal control – defining new designs for optimal biolocomotion system, and new principles for biolocomotive optimal control. They also illustrate a remarkable phenomenon. Energy-resonant states are not discrete, *e.g.*, restricted to particular frequencies, but in fact, are available in windows of frequency. The existence of these windows is due not to system nonlinearity, but to system damping – damping allows energetically-optimal frequency modulation.